# OPTIMISATION MODELS FOR TRANSPORTATION NETWORK DESIGN UNDER UNCERTAINTY: A LITERATURE REVIEW

Khadija Ait Mamoun[1, 2], Lamia Hammadi[1, 2],
Eduardo Souza De Cursi[2], Abdessamad El Ballouti[1]

[1]Laboratory of Engineering Sciences for Energy,
National School of Applied Sciences, UCD, El Jadida
[2] LMN, National Institute of Applied Sciences INSA
of Rouen, Saint-Etienne-du-Rouvray, France

*ABSTRACT*

*Supply chain network is critical to serving customers, so the most common practices are to determine the number, location, and capacity of facilities. But at the same time, uncertainties and risks must be taken into account in order to control delays. In this context, many optimisation models have been developed to use the results in transportation network and therefore improve the supply chain performance. Models were developed in both routing and zoning/districting problems, and different cases have been discussed in the literature, such as facility location problems, urban problems, and transportation problems. This paper seeks to review the literature in this area and decompose the models into Mathematical modelling and Geometrical approaches. Distribution is an important part of the supply chain management, it is a process with multiple participants. This characteristic brings a high level of uncertainty. This article therefore presents the distribution process and in particular the design of the transportation network which can include both routing and districting problems.*

## 1. INTRODUCTION

Nowadays, the complexity of routes generates uncertainties and hence, supply chain uncertainty becomes a concern for many researches and scientists [1, 2]. Delivery delays and quality problems are issues deriving from this complexity of supply chain network [3]. Due to this complexity, uncertainty in the supply chain is considered as an obstacle to the effective management and control of operations [4]. According to [5], "Supply chain uncertainty refers to decision making situations in the supply chain in which the decision maker does not know definitely what to decide as he is [unclear] about the objectives; lacks information about its environment or the supply chain; lacks information processing capacity; is unable to accurately predict the impact of possible control actions on supply chain behaviour; or, lacks effective control actions'. The importance of supply chain network appears in facility location decisions, and serving customers in good conditions.

The primary objective of the models presented in this paper is to find an optimisation solution in accordance with the problem. In this paper, the optimization approaches mentioned in the literature are divided into two classes: Mathematical modelling (e.g. Continuous approximation (Novaes et al .2009) & (Novaes et al .2004); Polynomial chaos expansion (PCE); The Monte Carlo Simulation (MCS) (Rafael Holdorf Lopez et al. 2010). ) and Geometrical models (e.g. Voronoi diagram, Visibility-shortest-path graph used by (Novaes et al .2009); ring-radial model used by (Novaes et al .2009) and (Novaes et al .2009) ) .

A large number of modelling philosophies such as expectation minimization, goal minimization deviation, maximum cost minimization, and the inclusion of probabilistic constraints have influenced the optimisation approaches under uncertainty [6]. In an optimisation process, the main techniques that take into account uncertainties are stochastic programming [7], fuzzy programming [8], reliability based design optimisation [9] and uncertainty quantification [10, 2, 11]. In engineering applications, robust design optimisation typically has the main objective of optimizing the mean performance of the system response and minimizing its variation. Besides the stochastic programming methods, Taguchi [12, 13] and other optimisation methods [14] have been applied to solve this kind of problem. Stochastic optimization and fuzzy optimisation differ mainly by the method of modelling the uncertainties. In the first one, the uncertain parameters are represented as random variables. In contrast, the latter considers such parameters as fuzzy numbers and constraints are treated as fuzzy sets [15].In engineering applications, the main objective of robust design optimization is usually to optimize the mean performance of the system response and the minimization of its variation [16].

Our framework concerns the distribution logistics process which has a direct impact on the supply chain performance, it is the link between production and the market. Hence, treating uncertainties problems into distribution process involves direct results on the supply chain process. Both routing and districting problems can be addressed as distribution logistics problems.

The vehicle routing problem (VRP) is a route optimization problem to minimize the total distance. The objective is to find optimal routes for multiple vehicles visiting a set of locations, with the aim of completing all deliveries as soon as possible. It is referred as the stochastic vehicle routing problem (SVRP) when some elements of the problem are random variables [17]. A review of the literature on the SVRP was provided by [18]. Due to Gendreau, a particular SVRP is modelled as a chance-constrained program (CCP) or as a stochastic program with recourse (SPR).The SVRP is treated by many authors using a mathematical programming formulation. On the other hand, some authors use approximate formulas to estimate the travel distances, rather than searching for specific optimal routes connecting the servicing points [17]. In addition, to solve some VRP other authors used a heuristics procedure [19].

On the other hand, districting problems are associated with a number of planning processes [17]. The objective of districting problems is to divide a certain area into smaller parts that we call districts or zones, taking into account some criteria such as balance, contiguity and compactness, and an objective function is optimised. Mathematical programming models are proposed to solve districting problems. Sections are grouped into a number of zones optimizing a convenient objective function under convenient constraints [21].

This paper is organized as follow: section 1 recalls the Literature authors and the main key words used in different papers and cases. We present a summary of the optimisation models used in the literature to solve the uncertainty problem in different case studies. Section 2 is dedicated to a classification of the optimisation models into two types: Mathematical modelling and Geometrical models. Section 3 discusses the case of distribution logistics, especially transportation process. In addition, the Routing and Districting problems are presented. Finally, we conclude in Section 4.

## 2. LITERATURE ANALYSIS: FACILITY LOCATION OPTIMISATIONS MODELS

Facility location optimization models are used recently by many authors and researchers, such as Antonio G.N. Novaes, Rafael Holdorf Lopez, Eduardo Cursi, etc, to find optimized solutions in different cases. Each author used the correct algorithm for his search taking into account the constraints and the criteria under consideration. Table 1 shows different algorithms used by the literature in different case studies related to Supply Chain Network uncertainty.

Table 1. Facility location optimisation models used in different case studies related to Supply Chain Network uncertainty

| The algorithm | The author who used the algorithm | Title | In which case the author have used the algorithm | Constraints and criteria of optimisation models |
| --- | --- | --- | --- | --- |
| Voronoi diagram : 1-Ordinary and multiplicatively weighted Voronoi diagrams 2-Power Voronoi diagrams | Antonio G.N. Novaes et al.2009 | Solving continuous location–districting problems with Voronoi diagrams | 1-Ordinary and multiplicatively weighted Voronoi diagrams are used to solve districting problems, and facility location problems. To minimize the maximum distance from a point to its nearest facility, the authors search for a set of facilities. 2- To solve districting problems with obstacles, Power Voronoi diagram is proposed. In this case, the obstacles are represented by two freeways located along the margins of two rivers, the Tietê and the Pinheiros. Those rivers are traversed by a number of bridges, which are congested most of the time. Hence, transportation operators do not design delivery zones covering two sides of the freeways. | *Balance*: To balance demand among districts *Contiguity*: The districts attained must be contiguous and compact. Contiguity is said to exist when we succeed to go from a point in the district to another point in the same district without having to pass through another district. *Compactness*: By analysing compactness, a penalty cost constraint is established that penalizes the "non-compactness" of the potential districts. |
| Voronoi diagram: The multiplicatively weighted Voronoi diagram | Antonio G.N. Novaes et al.2004 | A multiplicatively weighted Voronoi diagram approach to logistics districting | The multiplicatively weighted Voronoi diagram was used to smooth the contours of the district. The aim of districting in logistics distribution problem is to find an optimal partition to serve the delivery zones: this paper deals with delivery problems in the city of São Paulo, Brazil. | * Equality: the components which constitute a certain zone should lie inside a predetermined range. * The resulting parts or must be contiguous and compact. |
| Voronoi diagram: Planar Ordinary Voronoi Diagram ; Multiplicatively Weighted Voronoi Diagram ; Additively Weighted Voronoi Diagram ; Compoundly Weighted Voronoi Diagram | Antonio G.N. Novaes et al.2011 | Computational System to Determine the Optimal Bus-stop Spacing in order to Minimize the Travel Time of All Passengers | The authors used an approach by Voronoi diagram to determine the optimal bus-stop spacing which minimizes the travel time of all passengers, by combining this concept with non-linear programming. The bus line considered is located in the city of São Paulo, Brazil. | A big number of stops could be a good thing in that the user walks very little to arrive to the bus stop, but on the other hand, the route becomes tiring and distressing for those who travel a long distance. However, with a few stops the route is faster, but the users have to walk longer to arrive to the bus-stop. |
| Visibility-shortest-pathmetric | Antonio G.N. Novaes et al.2009 | Solving continuous location–districting problems with Voronoi diagrams | The visibility-shortest-path metric was used in association with Power Voronoi diagrams to solve districting problems with obstacles. | Obstacles in districting problems : for instance, represented by the mentioned rivers |
| Non-linearprogramming | Antonio G.N. Novaes et al.2011 | Development of a Computational System to Determine the Optimal Bus-stop Spacing in order to Minimize the Travel Time of All Passenger | Non-linear programming was used to find the ideal number of stops that optimizes the line for the users, by combining this method with Voronoi diagram. | Apply the optimal model for the number of stops. The fact that the distance between stops becomes a complicated process for the passengers who are accustomed to the original spacing. |
| The plane-sweep technique; the quad-tree technique and the continuous approximation | Antonio G.N. Novaes et al.2009 | Solving continuous location–districting problems with Voronoi diagrams | The authors used approximation algorithms associated with the construction of generalized Voronoi diagrams; which were used in applications to logistics and transportation problems. | Equalize the distribution effort among vehicles |
| Continuousdemand approximation | Antonio G.N. Novaes et al.2004 | A multiplicatively-weighted Voronoi diagram approach to logistics districting | The authors introduced a method to approximate continuous functions to spatial discrete variables such as the number of served points, the quantity of product to be distributed, and the stopping times. | * The number of points (customers) * The quantity of product to be distributed * The total stopping time consumed |
| Optimisation methods | Antonio G.N. Novaes et al.2009 | Solving continuous location–districting problems with Voronoi diagrams | The optimisation methods are used to solve districting and fleet design questions associated with logistics physical distribution problems. The aim of the optimisation models is to find the best vehicle size of a distribution problem with analysis. | The function is continuous but not well behaving, and therefore may not satisfy convexity assumption |
| Ring radial model | Antonio G.N. Novaes et al.2004 | A multiplicatively-weighted Voronoi diagram approach to logistics districting | To obtain an optimised solution, the radial ring model was used to the parcel distribution problem. It was used as a basis to apply MW-Voronoi diagram process. In other way, the radial ring districting was used as a starting pattern to determine the Voronoi diagram pattern. | Time and Capacity constraint: the aim of this optimisation model is to determine the best vehicle capacity and the best number of districts, respecting service demands. |
| Ring radial model | Antonio G.N. Novaes et al.2000 | A continuous approach to the design of physical distribution systems | The radial ring model was applied to the region that covers the city of São Paulo, Brazil. The partition of the area into districts was performed based on a polar coordinate system centered on a warehouse and using a dense radial ring road network, leading to wedge-shaped districts. | The number of visiting points assigned to each vehicle, will vary with cargo demands and stopping times, and should decline with the distance from the depot. |
| Polynomial chaos expansion (PCE) | Rafael Holdorf Lopez et al.2010 | Approximating the probability density function of the optimal point of an optimisation problem | The polynomial chaos expansion (PCE) was employed to minimize a stochastic functional. The optimised model could approximate the probability density function (PDF) of the global optimizer of one convex function, several non-convex functions and of a laminated composite optimisation, then the results of this methodology were compared to those of the Monte Carlo Simulation (MCS). | * Represent the random variables in terms of known random variables<br>* The construction of joint distributions which are generally unknown |
| The Monte Carlo Simulation (MCS) | Rafael Holdorf Lopez et al.2010 | Approximating the probability density function of the optimal point of an | The Monte Carlo Simulation (MCS) was used is employed in order to validate the results where different approaches were examined. In general, | The mean values of a certain number of samples are calculated. |

International Journal of Recent Advances in Mechanical Engineering (IJMECH)
Vol.11, No.1, February 2022| | | optimisation problem | the MCS are observed as *"true"* response that validate the results of the other approaches. | |
|---|---|---|---|---|
| Numerical optimisation of the stochastic functional | Rafael Holdorf Lopez et al.2010 | Approximating the probability density function of the optimal point of an optimisation problem | It is necessary to employ an optimisation method capable of handling noisy functions. In the case of convex functions, the algorithm is able to converge to the optimum point. For non-convex functions, a global optimisation algorithm must be introduced. | * The feasible set of the problem<br>* The projection operator |
| The Simulated Annealing (SA) algorithm | Novaes et al, 2016 | Process capability index Cpk for monitoring the thermal performance in the distribution of refrigerated products | The metaheuristic algorithm was used to optimise a refrigerated food delivery route: the aim is to minimise the distance travelled while taking into account the maintenance of the vehicle temperature and the capability level. | * Number of visiting points<br>* Quantity of product to be distributed<br>* Mean unloading time<br>* TTI (Time-Temperature Indicators ) values<br>* Process capability index |

We can conclude that an optimisation model is applied adequately to the situation presented by the author. In the majority of cases, the authors use more than one model combining different approaches in the same situation. In this context, we summarize the algorithms used in literature in table 2, in a way to indicate the Optimisation models used in different cases. Therefore, table 2 shows the summary of these models and their applications.

Table 2. Summary of algorithms used in literature depending on the case study

| The case study | Voronoi diagrams | Visibility-shortest path | Non-linear programming | The plane-sweep technique | The quad-tree technique | Continuous Approximation | Optimisation methods | Ring-radial model | PCE | MCS | Numerical optimisation of the stochastic functional | Simulated Annealing algorithm |
|---|---|---|---|---|---|---|---|---|---|---|---|---|
| **Districting problems** | X | X | X | | X | X | X | X | | | | |
| **Facility location problems** | X | X | X | | X | X | X | X | | | | |
| **Logistics Distribution problems and delivery problems** | X | | | | X | X | X | X | | | | |
| **Urban Transportation problem** | X | | X | | | | | | | | | |
| **Minimize stochastic functional. The** | | | | | | | | | X | X | | |

International Journal of Recent Advances in Mechanical Engineering (IJMECH)
Vol.11, No.1, February 2022| | | | |
|---|---|---|---|
| optimisation of non-convex unconstrained functions | | | |
| Treatment of noisy functions : Optimise convex and non-convex functions | | X X | |
| Refrigerated food delivery | | | X |

According to Table 2, the optimisation methods used to solve Supply Chain Network uncertainty are either mathematical or geometrical. Therefore, we can classify the approaches used for Facility Location Optimisation, into Mathematical modelling and Geometrical models.

## 3. LITERATURE FRAMEWORK FOR TRANSPORT NETWORK

In this section, we present the literature framework which represents different parts of our framework and which begins with a global scan of the literature until the specification of our study. Figure 1 represents the schematic of the framework:

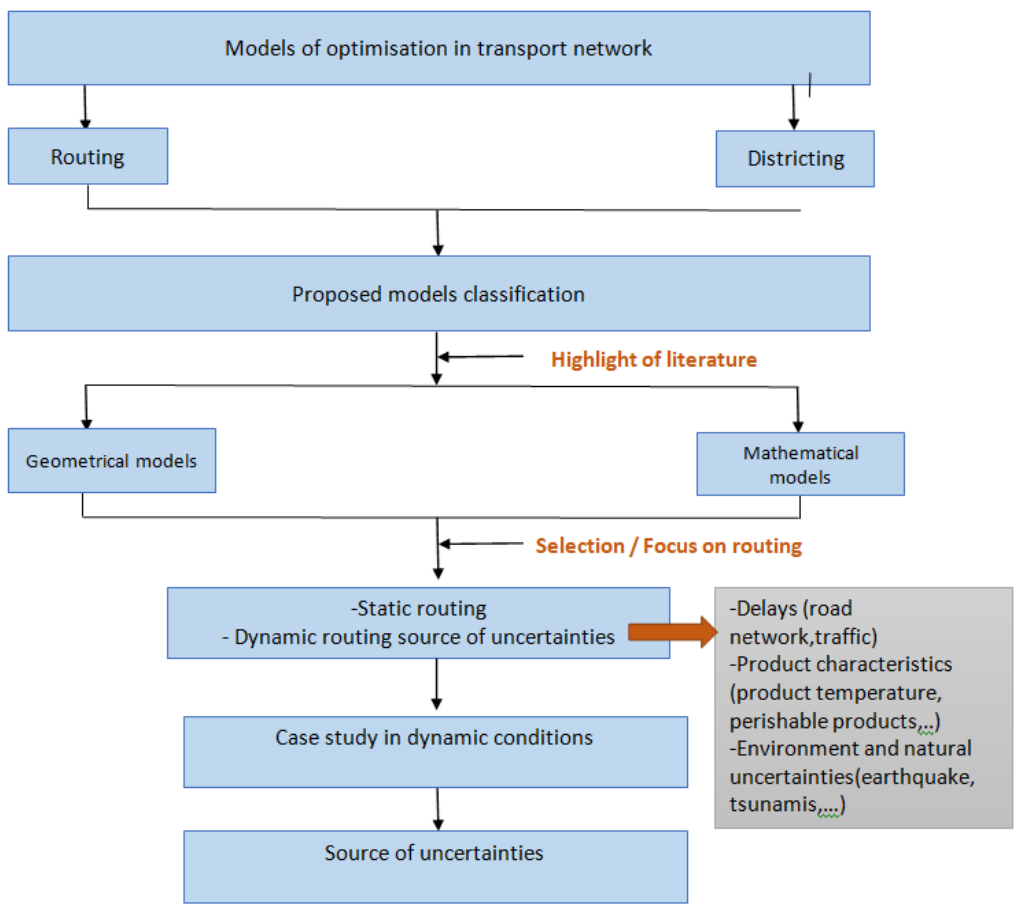

Fig 1. Literature framework for transport network

To achieve the objective of our case study, we have to use the adequate algorithm, and to do that, we propose a classification to differentiate the geometrical models from the mathematical models in order to adapt the correct algorithm to the situation.

## 4. CLASSIFICATION OF THE OPTIMISATION MODELS

Resulting from what was presented in the previous sections, we propose in this section a classification of the optimisation models in the transportation context into two categories Mathematical and Geometrical models. We suggest this classification in order to adapt the suitable model to the corresponding situation and to differentiate the contexts of the two approaches.

The figure 2 shows a classification of Optimisation models into Mathematical modelling and Geometrical models.

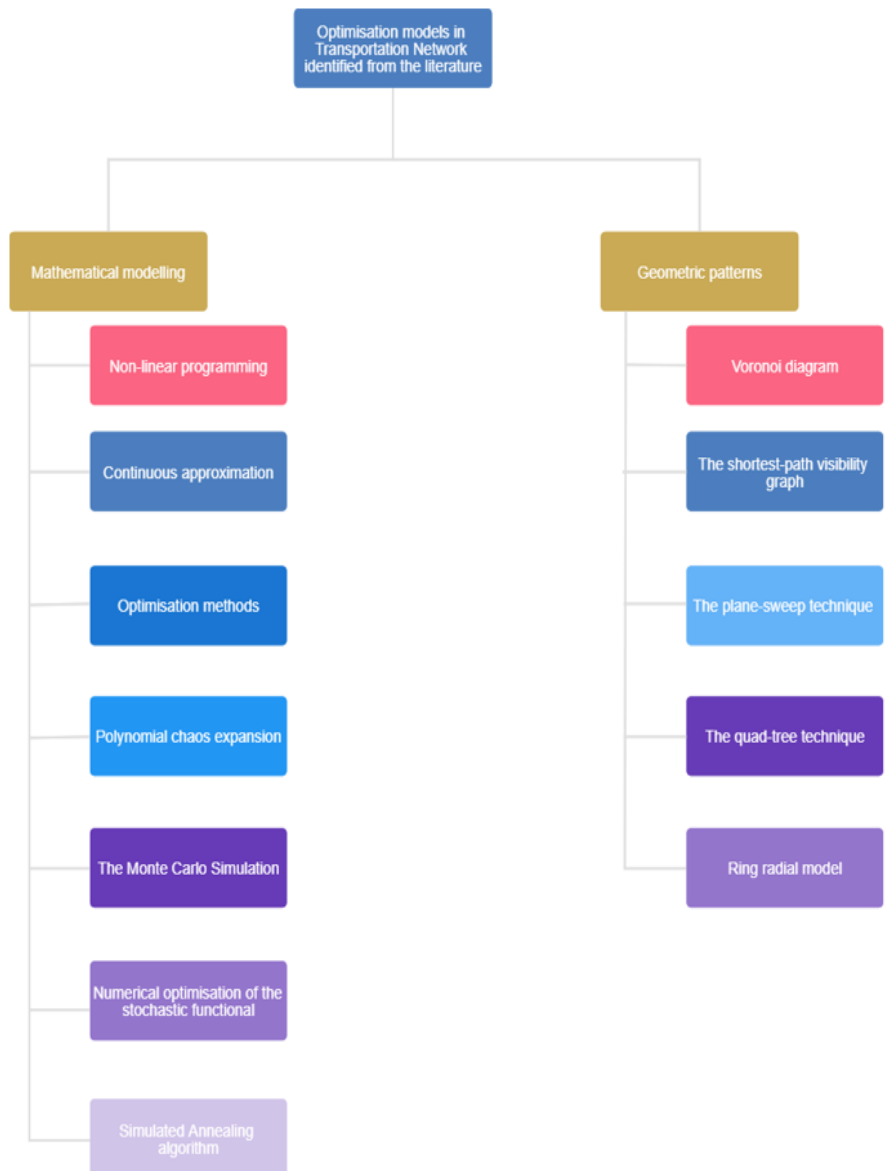

Fig 2. Classification of the optimisation models in transport network

## 5. UNCERTAINTY IN DISTRIBUTION LOGISTICS

Nowadays, the increasing flow of distribution as well as the high demand generates uncertainties and hazards that have a significant impact on the supply chain process, resulting in additional costs and delays. In this section, we will discuss the theoretical basis of some methods for solving transportation process problems. Two types of problems are considered: Routing and District problems.

### 5.1. Distribution logistics

Distribution logistics is considered as a connection between manufacturing and the market. Its domain of action includes all processes involved in distribution from manufacturing companies to customers. Customers are either end customers, distributors or processors. In concrete terms, distribution logistics includes goods handling, transport and interim storage [22].

The key objectives of Distribution logistics is : "The right objects must be available to customers in the right quantity at the right time with the right information at the right cost in the right place with the right quality" [22].

Distribution in supply chain management is a process where there are several participants. This characteristic leads to high uncertainty. Hence, the performance of the supply chain is closely related to the performance of the distribution process. It is important to decompose the distribution process in details in order to properly analyse uncertainty in distribution. Two major parts are presented: handling and transportation. Handling includes the procedures for loading and unloading goods, between the depot and the transportation vehicles. We can observe that handling is an internal procedure that can be better controlled than transportation, which is considered as an external procedure. Transportation can be impacted by many external factors, such as climate, traffic. Uncertainty in external environment is difficult to manage and is of increasing interest to scientists [23].

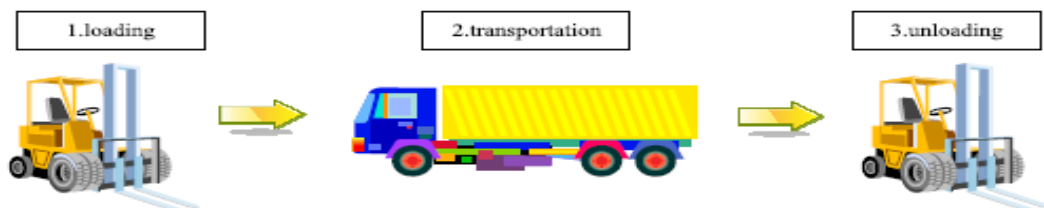

Fig 3. Distribution process

We can decompose the distribution process in details. The figure 3 shows the subdivision of distribution process: we can conclude that transportation plays a major role and that the handling part introduces less uncertainties than transport.

### 5.2. Transportation Network Design

Much of the literature focused on manufacturing operations. Nowadays, transportation uncertainties are increasingly affecting the performance of the supply chain. Therefore, we can conclude that: the Supply Chain Network, especially the transportation process is becoming more and more complex and the resulting uncertainties increase the risk and the vulnerability. This This has led many authors to research about risk and vulnerability in supply chain, such as [24].

The most important process in distribution logistics is the consolidation of the products, and delivery of distribution vehicles. Hence, the optimisation of the process of distribution logistics, especially transportation, becomes very important to achieve good results in terms of cost, speed and efficiency of the logistics transportation. So, two problems can be discussed: Routing and Districting problems.

### 5.2.1. Routing Problem in Transportation Process

In order to have the best delivery time and so satisfying the customer, we should implement a delivery planning in Transportation process with reasonable cost. In this kind of problem, we can distinguish two kinds of models, Traveling Salesman Problem (TSP) and Vehicle Routing Problem (VRP) [25].

**A-Traveling Salesman Problem (TSP)**

The beginning of the Traveling Salesman Problem dates back to the late 19th century: [26], and [27] used the first formal mathematical study of the problem [28]. The TSP has been solved by different approaches that have been evaluated in the literature: heuristic algorithms, meta-heuristic algorithms, such as Simulated Annealing, Threshold Accepting, Genetic Algorithms, Tabu Search, and Ant Colony Optimisation. Such a panel of methods is useful in a variety of optimisation problems. The TSP is based on the fact that a single vehicle visits a large number of points and then returns to the depot, while the total routing time or distance of the vehicle is minimized. Furthermore, when the vehicle capacity limitation is concerned this becomes a VRP.

**B- Vehicle Routing Problem (VRP)**

The Vehicle routing problem (VRP) is a problem where multiple vehicles are visiting a specific number of clients and the goal is to find optimal routes for a set of locations. The vehicles start from a central depot and go to deliver customers [29]. In other words, the use of vehicle routing (VRP) can involve minimum cost routes while serving customers with known demands starting and terminating at a central depot. Each customer is served exactly once and all the customers must be allocated to vehicles such that the vehicle capacities are not exceeded [30].

The VRP was presented and solved by literature by different approaches: exact algorithms, heuristic methods, etc.

**C- Comparison between TSP and VRP**

The difference between a VRP and a TSP is that in VRP multiple vehicles are visiting a set of points and numerous routes are produced, but when there are no constraints, we talk about a TSP where only one vehicle is visiting all locations while finding the shortest route. The difference between TSP and VRP is illustrated in Figure 4.

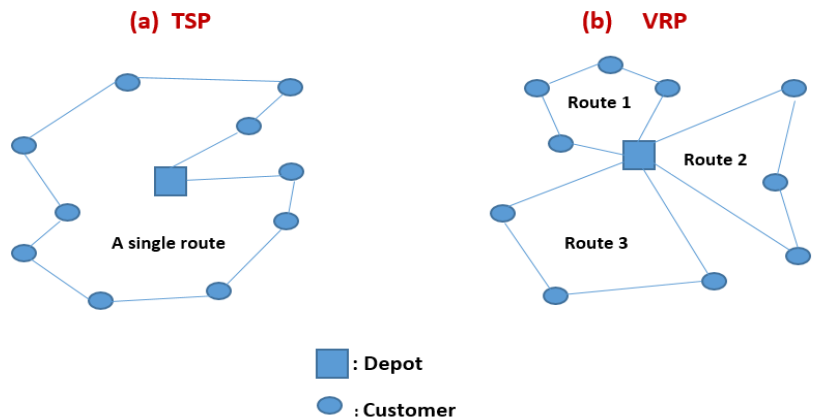

Fig 4. Comparison between TSP and VRP. Source Wan-Yu Liu et al.2014

**I) Districting Problem in Transportation Process**

Multiple case studies of the districting problem have been presented through literature such as: political districting [31], police zone districting [32], districting for salt-spreading operations [33], school districting [34], homecare zone design [35] electric zone design [36],..etc. During these applications, multiple objective functions and constraints appeared. In consequence to this, a number of heuristics were designed depending on the districting problem situations.

There are other terms for district that can be cited such are territory, sector, zone, region, or area . These sub-divisions usually follow some constraints. For example, there are small geographic areas that are indivisible, so-called basic areas. District can be defined as a composition of basic areas.

We can cite typical examples for basic areas such as cities, zip-code areas, streets, and single customer locations.

According to [33], districting related with transportation problems should be performed at the strategic level, while routing should be performed at the operational level. In other words, districting call for a global vision and is often related to the managerial and administrative levels, while routing processes are more detailed and linked to day-to-day operations. The main constraints of the districting problems:

- *Contiguity* : we talk about contiguity when it is possible to move from a point in the district to another point in the same districting without passing through any other district.
- *Compactness*: we talk about compactness when the district is closely and nearly packed in an undistorted
  shape : it is suggested that districts be circular or square rather than elongated.
- *Balance or equity*: Districts are balanced when they have approximately the same size in a quantifiable operation such as number of inhabitants, sales potential, or working time.
- *Respect of natural boundaries*: Rivers, railroads, mountains, administrative boundaries, etc.
- *Socio-economic homogeneity*: represent a good representation of population who share the same interests and views.

### A- *Components of districting problem*

- Basic Areas: In districting or zoning problems, basic areas are also calledbasic units, coverage areas, or geographical units. A basic area is the smallest geographic area and it is represented by a point.
- Centers: Each district can be associated to a center. This can either be a specified location, e.g., an office of a company or a location of a social institution, or a representative point, e.g., the center of gravity.
- Distances : The distance between two basic areas
- Districts: A district consists of a set of basic areas. A district can be called zone, territory or sector.
- Districting Plan: Districting plan represents a set of districts. Districting plan can be called districting layout, territory plan, territory layout or solution.

### B-*Methods to solve Districting problems*

*(1) Mathematical modelling*: In the literature, some approaches model the districting problem as a mathematical program. [37] proposed the first formulation as mixed integer program. This program considers compactness as an optimisation objective, while balance is treated as a hard criterion. The model does not contain any contiguity constraints. Someresearchers have added contiguity constraints to the early models especially Hess' model, which does not model contiguity explicitly. Contiguity was introduced by[38]. Each basic units excepting the center, is a source of one supply, while the center contains multiple demands.

*(2) Heuristic Solution Approaches*

   - *Location-Allocation:*

Originally, Hess used a location-allocation heuristic and divided the problem into two different ones. Already Hess has proposed a location-allocation heuristic that splits the problem into two independent problems. The location problem consists of determining a set of centers, while the allocation problem consists of determining when base units are assigned to these centers. The two problems are treated alternately until no further improvement is possible. One way to solve the location problem is to determine the center of gravity for each area in the previous allocation phase. Many authors have applied the location-allocation procedure over the years.

   - Further approaches: Many other heuristics have been proposed over the years. For example,[39]presents a set-partitioning approach. Basically, this approach form a set of feasible districts. Then, a subset of these districts is chosen to obtain a good general solution. Some base areas are chosen by seed growth approaches as seeds and assign the other base areas to these seeds while maintaining the required planning criteria, each seed leads to a district. The districts are treated either sequentially or simultaneously. Many meta-heuristics have been proposed in the context of districting. The characteristic of these metaheuristics is that they are very flexible to integrate different requirements and planning criteria. Many other heuristics have been proposed over the years. For example, [39] presents an ensemble partitioning approach. Basically, this approach forms a set of feasible districts.

(3) Geometrical Approaches: The "Recursive Partitioning Algorithm (RPA) "is based on the work of[40]. It employs the principal geographical information of the districting problem.Another geometrical model to solve districting problems are discussed previously in the Chapter 4 which are Voronoi Diagrams, The shortest-path visibility graph, The ring-radial model, The plane-sweep technique and the quad-tree technique.

## II) CASE STUDY FORMULATION

This framework starts with a review of the literature on optimization models, and then a classification of these models into mathematical modelling and geometrical models The models used in the literature have been implemented in different cases of transportation networks: [17] presents an application of the radial ring model in the case of urban distribution, [41] takes the example of public transportation. [42] optimizes the delivery of refrigerated food products with the simulated annealing algorithm, taking into account the temperature variation and its impact on the product quality. The models used in the literature can be classified into mathematical modeling and geometric models. In our framework, we present the distribution process, especially the transport part. Hence, we can call for vehicle routing problem. Due to the increase in road traffic, the uncertainties related to transportation directly affect the performance of the supply chain. Different factors and sources of uncertainties are identified in the transportation domain: (e.g. product characteristics, process/manufacturing, control/response uncertainty, decision complexity, customer demand, environment, natural uncertainty,...). Thus, when there is uncertainty in transportation, it means that the framework and research must be studied under dynamic conditions. In this situation, the information is unknown or uncertain at the time of planning. The information may change after the initial route has been constructed. In this type of setting, we use the Dynamic Vehicle Routing Problem (DVRP). In DVRP, various elements causing dynamism can be identified such as:

- Dynamic consumers: dynamic requests, dynamic localisation, dynamic demands, dynamic customer availability, dynamic order time.
- Dynamic travel times: weather or traffic conditions delay arrival at the given customer.
- Dynamic service time: the planning of adjustment time can change due to the problems identified on the site.
- Dynamic vehicle availability: unexpected events like vehicle breakdowns.

In addition to the constraints mentioned above, there is also that of temperature variation, in particular cases of distribution such as the delivery of refrigerated food products, the distribution of pharmaceutical products, perishable foodstuffs... In this type of situation, the quality of the product can change due to temperature variations during the distribution process. To consider the temperature criteria, researchers have used the process capacity index, [41] used the six-sigma methodology to quantify the performance of refrigerated food distribution.. In general, the capability indices are considered as a link between the process parameters and engineering specifications. Thus, the contribution of our framework is to optimize a route distribution under dynamic conditions using a metaheuristic algorithm considering the temperature constraint. Figure 4 presents the methodology of our framework approach.

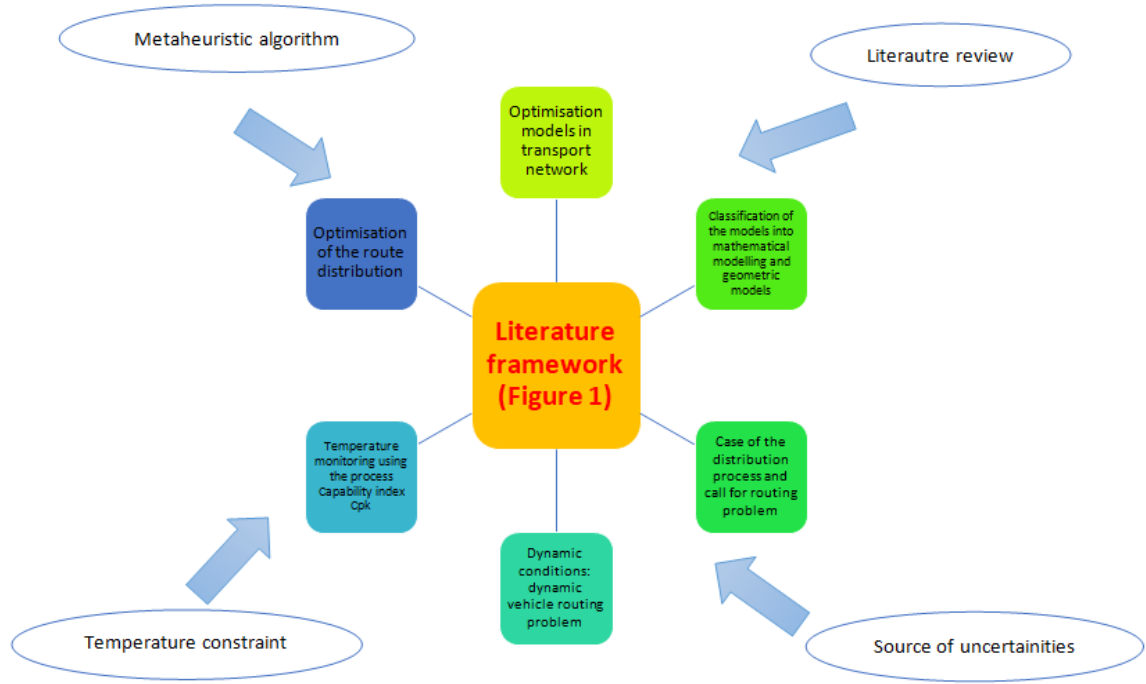

Fig 5. The case study formulation

## THE CASE STUDY

The distribution process is an important step in the pharmaceutical industry, especially as human health is increasingly attracting the world's attention. Therefore, to analyze the quality of the product, the process capability index (Cp, Cpk, Cpm...) is used to measure the process performance [43]. The authors of the study [44] used process capability indices to quantify process performance in the pharmaceutical industry. When product quality is related to a multitude of characteristics, we use "multivariate process capability indices"[45]. Our framework will take into account the temperature constraint during the dispensing process under dynamic conditions, using a metaheuristic algorithm such as the "simulated annealing algorithm".

## III)   CONCLUSION AND FUTURE WORK

The objective of this research review is to present the different optimization models used in the literature. The authors have used the optimization models in different case studies to solve different transportation network problems. We divided the optimization models into mathematical and geometric models and some examples were presented. Our framework is based on the distribution process and more specifically on transportation, which has a direct impact on the performance of the supply chain. In this context, routing and zoning/districting problems are addressed. The literature has developed many solutions to these types of problems by combining operations research methods and computational techniques to obtain practical solutions.

In the transportation process, uncertainty problems are becoming more and more complex with the increase of traffic. In this context, our future framework will focus on the Dynamic Vehicle Routing problem to obtain an efficient routing solution under uncertainty. In this approach, not all information is known in advance, such as the delivery or repair of an equipment. The objective of the future framework is to define the relationship between product quality and distribution and to evaluate the impact on vehicle routing problems.

**ACKNOWLEDGEMENTS**

This work is supported by Toubkal Project/22/143, Campus France and CNRST Morocco

## AUTHORS

**Khadija Ait Mamoun** is a PhD student at industrial and logistic engineering of the ENSA of El Jadida, Energy Engineering Science Laboratory (LabSIPE), under the theme: "Quantification of uncertainties in the design of transport networks". Scientific fields: Transport, Logistics, Applied Mathematics.
Has an engineering degree in industrial engineering and production in 2010 from the « International Academy Mohammed VI of Civil Aviation». Her research areas include: Logistics, Transport Engineering, Supply Chain Management, Risk Analysis and Management, Uncertainty Quantification, Mathematical modelling, Optimization, Operational research.

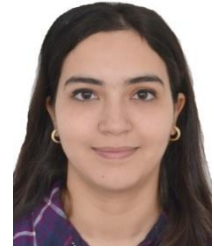

**Lamia. Hammadi**, is Assistant Professor of Industrial and Logistics Engineering at National School of Applied Sciences, ChouaïbDoukkali University of El Jadida, Morocco. Lamia graduated as PhD in Industrial and Logistics Engineering under the theme "Customs supply chain Engineering: Modelling and risk management: application to the customs" from National School of Applied Sciences, Cadi Ayyad University of Marrakech, Morocco and National institute of Applied Sciences of Rouen, France. She graduated as an Industrial Engineer from National School of Applied Sciences of Fes, Morocco, in July 2011.Her research areas include Logistics engineering, Supply chain Management, Stochastic Modelling, Uncertainty Quantification, Facility location, Risk Analysis, Risk Management.

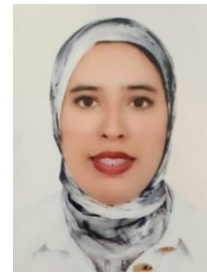

**Abdessamad El Ballouti** is Qualified Professor at National School of Applied Sciences ,ChouaïbDoukkali University of El Jadida, Morocco. A. El Ballouti graduated as Qualified Professor at Energetic Engineering in 2016. A. El Ballouti graduated as PhD in Electronic Engineering in 2010 from CHouaibDoukkali University.
Her research areas include: Energetic, Electonic, Materials, Logistics, Transport Engineering, Risk modeling.

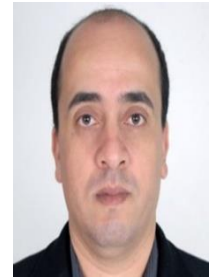

**Eduardo Souza de Cursi** graduated in Physics from the University of São Paulo, master's degree in physics from the Brazilian Center for Physics Research and Docteuren Sciences - Université des Sciences et Techniques du Languedoc. It is a full professor at the Institute National des Sciences Appliquées of Rouen, France. He is currently Director of the Laboratoire de Mécanique de Rouen, Director of European Affairs and International of INSA Rouen and responsible for the Franco-Dominican training program Dominicans engineers in France. He has experience in Applied Mathematics and Theoretical Mechanics area, with emphasis on Numerical Analysis, Stochastic Methods and Convex Analysis.

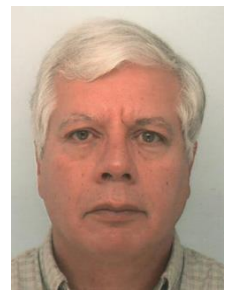